\documentclass[11pt]{article}
\pdfoutput=1

\usepackage[utf8]{inputenc}
\usepackage[T1]{fontenc}
\usepackage{amsmath,amssymb}
\usepackage{graphicx}
\usepackage{booktabs}
\usepackage[skip=8pt,labelfont=bf]{caption}
\usepackage[hidelinks]{hyperref}
\usepackage[round]{natbib}
\usepackage{geometry}
\title{Measuring the engine of a liquidation cascade:\\
subcritical branching inside a first-order transition}

\author{Ramon Marc Garcia Seuma\thanks{Correspondence: reymon.devs@gmail.com}}
\date{\today}

\begin{document}
\maketitle

\begin{abstract}
We study seven major crypto-perpetual liquidation cascades (2022--2025), and in
the largest of them we can watch the mechanism directly. From the on-chain fill
log of a fully transparent venue we measure the branching ratio of that
event---the October~2025 crash, the largest on record---\emph{in flight}, with
both of its factors observed and no free constants. It ran deeply subcritical: the structural ratio and the
amplification bookkeeping both place it at $\hat\lambda \approx 0.1$--$0.2$
throughout, and a third, flow-based estimator---whose calm-market level is
mechanically inflated and must be read as a trajectory---\emph{falls} through the
climax rather than rising. All three agree on subcriticality within the venue,
not on a common numerical level. Alongside them, $88\%$ of all post-onset forced
selling landed within thirty minutes and $63\%$ of it was absorbed off-book by the
venue's backstop, which drives the branching ratio \emph{down} precisely at the
climax. Across the full set of seven, at onset---the minute ending the steepest
hour of each crash---the order parameter (mean inter-asset coupling) jumps by
between $1.6$ and $4.4$ baseline standard deviations into a near-fully-ordered
phase, while the susceptibility proxy $\chi = N\,\mathrm{Var}(c_{ij})$
\emph{collapses} in five of the seven events and diverges in none; the jump is
invariant under subsampling ($N = 8$--$28$). The transition is abrupt and
scale-robust rather than critical, and its in-cascade signature lives in the
liquidity
sector: price impact spikes on two venues and two instruments (a factor of
$1.2$--$3.5$ regressed, $3.2$--$9.1$ quoted directly) while open interest clears
by $25$--$70\%$. The natural mechanistic account---a Galton--Watson cascade with
$\lambda = k\tilde\rho$ (impact per forced dollar $\times$ forced notional per
unit relative move)---is then eliminated as a description of the
\emph{pre-cascade state}: both of its falsifiable predictions fail at simulated
power $\ge 0.96$, on proxied and on directly measured regressors alike. Severity
is set by shock $\times$ map-in-path $\times$ liquidity withdrawal rather than by
a diverging multiplier, which is why none of the scalar pre-state measures we
can construct grades it.
\end{abstract}

% ---------------------------------------------------------------------
\section{Introduction}
\label{sec:intro}
Whether large market crashes are \emph{critical transitions}---endogenous
instabilities that a system approaches gradually, shedding resilience until an
infinitesimal shock tips it over---is among the oldest questions in the physics
of markets. The program has a long lineage, from the log-periodic critical-point
picture of \citet{johansen2000crashes} and \citet{sornette2003why} to its
operational arm, the theory of early-warning signals (EWS): near a critical point
the dominant relaxation mode slows, so fluctuations grow more autocorrelated and
more variable ahead of the transition \citep{scheffer2009early}. Part~I of this
series \citep{garcia2026ews} put that operational claim to the test on seven major
BTC perpetual-futures liquidation cascades (2022--2025) and found it wanting: no
single-variable early-warning signal is event-invariant, and the failures are
structured---critical slowing down is absent precisely where the shock is most
abrupt, echoing earlier doubts that financial meltdowns are critical transitions
at all \citep{guttal2016lack}.

A negative result on \emph{indicators} poses, but cannot answer, the prior
question: what is the \emph{order} of the transition? Criticality is a statement
about collective structure, not about any single series, and a first-order
transition can jump discontinuously with no variable slowing beforehand---exactly
the signature Part~I reported. Order must therefore be measured where it lives, in
the market's correlation fabric, rather than inferred from the memory of one
price. Section~\ref{sec:fabric} does this directly.

A leveraged market makes a second, usually inaccessible, question answerable: the
\emph{mechanism}. The feedback laws here are not hidden model assumptions but
published exchange rules---the liquidation engine, funding, and the backstop that
clears insolvent positions. The natural mechanistic account is a branching
cascade, in which each forced sale moves the price and triggers, on average, some
number of further liquidations \citep{thurner2012leverage,filimonov2012quantifying,cont2016fire}.
On Hyperliquid, a fully on-chain venue, every open position and every forced fill
is public, so the offspring mean of that cascade---the branching ratio
$\lambda$---can be \emph{measured} rather than fitted from aggregate reflexivity.
We do exactly that, for the October~2025 cascade---the largest single-day
liquidation event recorded in crypto-perpetual markets
\citep{amberdata2025oct,chitra2025adl}---and the answer is that it never came
near the critical boundary.

We make four contributions, listed here in reading order. (i) We characterize the
transition as \emph{first-order in its signatures}: at onset the order parameter
jumps into a near-fully-ordered phase while the susceptibility proxy
\emph{collapses} rather than diverges, and the jump is invariant under
subsampling---a control that rules out panel composition and small-panel
artefacts as its source (Sec.~\ref{sec:fabric}). (ii) We locate the in-cascade
signature in the liquidity sector: price impact spikes and open interest clears,
on two venues and two independent measurements of impact (Sec.~\ref{sec:liquidity}).
(iii) We eliminate the branching model \emph{as an account of the pre-cascade
state}: both of its falsifiable predictions fail with simulated power $\ge 0.96$,
on proxied and on directly measured regressors alike (Sec.~\ref{sec:lambda}).
(iv) And---the result that motivates the other three---we provide, to our
knowledge, the first \emph{in-flight} measurement of a record cascade's branching
ratio, from Hyperliquid's fill log, with both factors observed and no free
constants: subcritical throughout, front-loaded almost to instantaneity, and with
most of the offspring absorbed off-book by the venue backstop, so that the
exchange's own liquidation waterfall suppresses the very feedback the model is
built on---a mechanism-design result with a falsifiable cross-venue prediction
(Sec.~\ref{sec:engine}). Read together, (iv) explains (iii): a cascade whose
multiplier never approaches unity has no diverging quantity for a pre-state
variable to track.

The remainder proceeds through data (Sec.~\ref{sec:data}), the four results, and
a synthesis (Sec.~\ref{sec:discussion}). The results are presented as a build
rather than in order of importance: the collective characterization first
(Sec.~\ref{sec:fabric}--\ref{sec:liquidity}), then the branching model and its
elimination (Sec.~\ref{sec:lambda}), which is where $\lambda$ is defined, and only
then the in-flight measurement of that same quantity (Sec.~\ref{sec:engine}).
Two contemporaneous works frame ours:
\citet{shukla2026networks} build high-frequency crypto interaction networks in
calm markets, a useful contrast to the in-cascade fabric studied here, and a
parallel preprint \citep{rs9459584} compares CeFi and DeFi microstructure through
the same October~2025 event. We cite both as concurrent rather than antecedent.

% ---------------------------------------------------------------------
\section{Data}
\label{sec:data}
Our analysis draws on three sources, each with boundaries stated at the outset.
The first is a panel of the most liquid Binance USDT-margined perpetuals sampled
at five-minute resolution across seven event windows; the panel grows from $N=22$
names in May~2022 to $N=30$ in 2024--2025, a composition change we control for
explicitly with the finite-size analysis of Sec.~\ref{sec:fabric}. The second is
the Binance BTC series---one-minute klines and five-minute derivatives
metrics---which underpins the impact regressions and the severity designs. The
third is Hyperliquid: a per-minute archive of asset contexts, whose quoted impact
prices furnish a \emph{direct} measure of price impact $k$ (no regression
required) and of open interest across five events, and the venue's on-chain fill
log, which attributes every forced fill to a liquidated user. The fill log is
decisive but time-bounded: it begins on 2025-05-25, so a severity test on
\emph{measured} liquidation density across the historical events is structurally
impossible---the October~2025 cascade is necessarily the single case-study arm for
the in-flight measurement (Sec.~\ref{sec:engine}). One methodological caution
recurs: in the fill log the liquidation object rides both legs of each trade, so
all totals are deduplicated by the liquidated-user leg or they double-count.

\paragraph{Events and the onset convention.}
Every quantity in this paper is measured relative to a single per-event
reference time, the \emph{onset}. We define it mechanically: the onset is the
minute that terminates the most negative 60-minute log return of the event,
located automatically inside a search window pinned to the documented cascade
date. The search window is necessary rather than cosmetic---run unrestricted
over a two-month file, the detector selects a different, larger drawdown inside
the same file for three of the seven events. Fixing the onset this way makes
``pre-onset'', ``days since onset'' and ``the jump at onset'' mean the same
thing across events with very different durations, and it commits the
convention before any statistic is computed. Table~\ref{tab:events} lists the
seven cascades, their onsets, and the endogenous/exogenous typing we carry
through the paper. That typing is \emph{empirical, not causal}: it labels the
two events in which Part~I found no critical slowing down in price
autocorrelation (Feb~2025, Oct~2025) as \emph{exo} and the remaining five as
\emph{endo}. Both exo events happen to be abrupt tariff-news shocks, which is
the reading that motivated the label, but the correspondence is not clean---the
April~2025 tariff cascade carries a news trigger of the same kind and still
shows the endogenous price signature. We therefore treat the typing as a
hypothesis inherited from Part~I and a convenient grouping for the figures,
never as a validated taxonomy, and no claim in this paper rests on it.

\begin{table}[t]
  \centering
  \caption{The seven liquidation cascades. Onsets are UTC, auto-detected as the
  minute ending the most negative 60-minute log return within a window pinned to
  the documented event. Type is the empirical grouping inherited from Part~I
  (\emph{exo} = no critical slowing down in price autocorrelation), not a causal
  classification by trigger; see the text. $N$ is the size of the perpetuals panel
  available in that era. Source: EXP-006.}
  \label{tab:events}
  \begin{tabular}{lllcc}
    \toprule
    label & trigger & onset (UTC) & type & $N$ \\
    \midrule
    May '22 & LUNA/UST collapse            & 2022-05-11 12:51 & endo & 22 \\
    Nov '22 & FTX insolvency               & 2022-11-08 19:32 & endo & 25 \\
    Aug '24 & yen carry-trade unwind       & 2024-08-05 01:10 & endo & 30 \\
    Dec '24 & December leverage flush      & 2024-12-09 15:56 & endo & 30 \\
    Feb '25 & first tariff announcement    & 2025-02-03 02:07 & exo  & 30 \\
    Apr '25 & ``Liberation Day'' tariffs   & 2025-04-07 15:17 & endo & 30 \\
    Oct '25 & record liquidation cascade   & 2025-10-10 20:50 & exo  & 30 \\
    \bottomrule
  \end{tabular}
\end{table}

% ---------------------------------------------------------------------
\section{The transition is first-order in its signatures}
\label{sec:fabric}
A word on what we claim. ``First-order'' here is an \emph{empirical
classification by signature}, not a thermodynamic one: we do not take an
$N\to\infty$ limit, control a temperature, or construct a free energy. The
operational criterion is the conjunction of three measurable facts---a
discontinuous jump in the order parameter, no divergence of the susceptibility
proxy through the transition, and persistence of both under panel
subsampling---and what it licenses is the rejection of the \emph{critical-point}
reading in this proxy over these windows, not a proof that criticality is absent
from leveraged markets in general. A poorly chosen proxy or a strongly
non-equilibrium system could hide a signature that is really there. The
contribution is making the discrepancy measurable. The classification is also
\emph{recurrent rather than universal}: the sample is heterogeneous, and the two
exceptions are named throughout---May~2022, a multi-day grind whose fabric was
already ordered and which therefore has no discrete jump to make, and
October~2025, whose $\chi$ does not collapse across its own cascade.

We measure the transition in the cross-asset correlation structure. The order
parameter is the mean pairwise coupling $\bar c = \langle c_{ij}\rangle$
(equivalently, the weight of the leading market mode); the susceptibility proxy is
$\chi = N\,\mathrm{Var}(c_{ij})$, the extensive variance of couplings that would
diverge at a genuine critical point.

Across the seven events the order parameter jumps at onset by $+1.6$ to $+4.4$
base standard deviations in six of seven cases, into a near-fully-ordered phase
with $\bar c \approx 0.85$--$0.90$ and leading-eigenvalue weight
$\lambda_1/N \approx 0.9$ that persists for days; the sole exception is the
May~2022 grind, whose fabric was already ordered before onset. The susceptibility
does the opposite of what criticality requires: $\chi$ \emph{collapses} through the
transition---a negative jump in five of seven events, reaching $-3.4$~sd---and
diverges in none. The minimum spanning tree contracts abruptly; for October~2025
its normalized length falls from $0.60$ to $0.24$ in a single step, with a second
collective mode emerging above the Mar\v{c}enko--Pastur edge of the random-matrix
spectrum \citep{laloux1999,plerou2002} at the cascade. That last criterion is
conservative in exactly this regime, and asymmetrically so. The nominal edge
$(1+\sqrt{N/T})^2 \approx 1.75$ (here $N = 30$, $T = 288$) is fixed at unit noise
scale, whereas in the ordered phase the market mode absorbs most of the trace: with
$\lambda_1/N \approx 0.89$ the residual bulk carries variance
$\sigma^2_{\mathrm{res}} = (N-\lambda_1)/(N-1) \approx 0.11$, so an edge
recalibrated on the deflated spectrum sits near
$\sigma^2_{\mathrm{res}}\,(1+\sqrt{(N-1)/T})^2 \approx 0.19$. A second eigenvalue
clearing the nominal edge in that regime therefore clears the recalibrated one by
roughly a factor of nine. The
corollary is that the criterion must not be read in the opposite direction---a
\emph{decline} in the count of supercritical modes through the transition is
induced mechanically by the growth of $\lambda_1$ and is not evidence that
subdominant factors merge. Separating the two requires an adaptive edge
recalibrated on the deflated spectrum together with an eigenvector delocalization
filter \citep{garciamedina2026igf}; we do not attempt it here.

The decisive control is finite size. A discontinuous onset jump could in principle
be a crossover artifact of a small, growing panel; the subsampling analysis rules
this out. The susceptibility is extensive in every event and regime (for
December~2024 the baseline $\chi$ rises from $0.18$ to $0.77$ as $N$ grows from $8$
to $28$), yet the \emph{onset jump is $N$-invariant}: across the subsampling
range---now expressed in base-regime sd units ($[-10,-3)$~d, which exclude the
pre-onset build-up and therefore read somewhat larger than the pre-10-day units
quoted above)---it holds at $3.1$--$3.3$ (April), $5.5$--$5.65$ (December), $2.58$--$2.66$
(August), $2.26$--$2.33$ (November), $1.64$--$1.66$ (February), and $1.67$--$1.81$
(October), and the ordered-phase collapse of $\chi$ persists at every $N$. A
scale-invariant jump with a collapsing, extensive susceptibility is the signature
of a first-order transition, not a critical point.

Structure and single-variable precursors dissociate. February~2025, which carried
\emph{no} single-variable early-warning signal in Part~I, shows the
\emph{strongest} fabric build-up. We summarise a pre-onset trend by the
Kendall rank correlation $\tau$ between an observable and time over the three
days before onset, so $\tau > 0$ means the quantity drifts upward into the
event and $|\tau|$ measures how monotone that drift is; for February~2025 the
mean coupling gives $\tau = +0.83$ ahead of onset --- the only pre-onset trend
that survives block resampling on both fabric metrics ($p \le 0.006$;
Table~\ref{tab:fabric}). October~2025 is the fabric outlier in the
opposite direction, de-correlating into onset (pre-onset $-0.50$, $\chi$ jumping
$+0.47$; descriptive under the resampled null) even as it is the most violent
event by every liquidity measure below. The order of the transition is thus not
readable from any one series---the motivating fact of Part~I, now given its
structural cause.

Two boundaries are load-bearing. The event windows overlap (adjacent estimates
share $11/12$ of their data), so nominal trend significances are optimistic; we
re-derive them with a circular moving-block bootstrap against each event's full
pre-onset stretch (block $= 12$ steps $=$ the one-day window length, $B=5000$).
Under that null most pre-onset trends are \emph{not} significant
(Table~\ref{tab:fabric}): only the February~2025 build-up and the November~2022
order-parameter drift ($p = 0.009$) survive. The onset jumps---the load-bearing
result---are regime means, not trend statistics, and are unaffected. And the
coupling is a linear correlation proxy, which understates tail co-movement.
Neither boundary affects the qualitative result: a jump-with-collapse repeated
across seven events and stable under subsampling.

\begin{table}[t]
  \centering
  \caption{The fabric transition per event: onset jumps (first post-onset day
  vs.\ the 3-day pre-onset mean, in pre-10-day sd units) and pre-onset 3-day
  Kendall trends with block-resampled two-sided $p$-values (circular
  moving-block bootstrap on each event's full pre-onset stretch; block $= 12$
  steps $=$ the one-day window length; $B = 5000$). Bold: $p < 0.05$.
  Sources: EXP-012, EXP-020.}
  \label{tab:fabric}
  \begin{tabular}{llcccccc}
    \toprule
    & & \multicolumn{2}{c}{onset jump} &
    \multicolumn{2}{c}{pre-onset trend in $\bar c$} &
    \multicolumn{2}{c}{pre-onset trend in $\chi$} \\
    \cmidrule(lr){3-4}\cmidrule(lr){5-6}\cmidrule(lr){7-8}
    event & type & $\Delta\bar c$ [sd] & $\Delta\chi$ [sd] &
    $\tau$ & $p$ & $\tau$ & $p$ \\
    \midrule
    May '22 & endo & $-0.08$ & $+0.30$ & $+0.43$ & $0.263$ & $-0.62$ & $0.062$ \\
    Nov '22 & endo & $+1.90$ & $-0.33$ & $+0.81$ & $\mathbf{0.009}$ & $-0.58$ & $0.107$ \\
    Aug '24 & endo & $+2.43$ & $-1.92$ & $-0.08$ & $0.874$ & $+0.33$ & $0.434$ \\
    Dec '24 & endo & $+4.41$ & $-3.40$ & $+0.40$ & $0.350$ & $+0.26$ & $0.564$ \\
    Feb '25 & exo & $+1.67$ & $-0.88$ & $+0.83$ & $\mathbf{0.006}$ & $-0.83$ & $\mathbf{0.006}$ \\
    Apr '25 & endo & $+2.68$ & $-2.45$ & $+0.17$ & $0.731$ & $-0.63$ & $0.082$ \\
    Oct '25 & exo & $+1.56$ & $+0.22$ & $-0.50$ & $0.202$ & $+0.47$ & $0.250$ \\
    \bottomrule
  \end{tabular}
\end{table}

\begin{figure}[t]
  \centering
  \includegraphics[width=\linewidth]{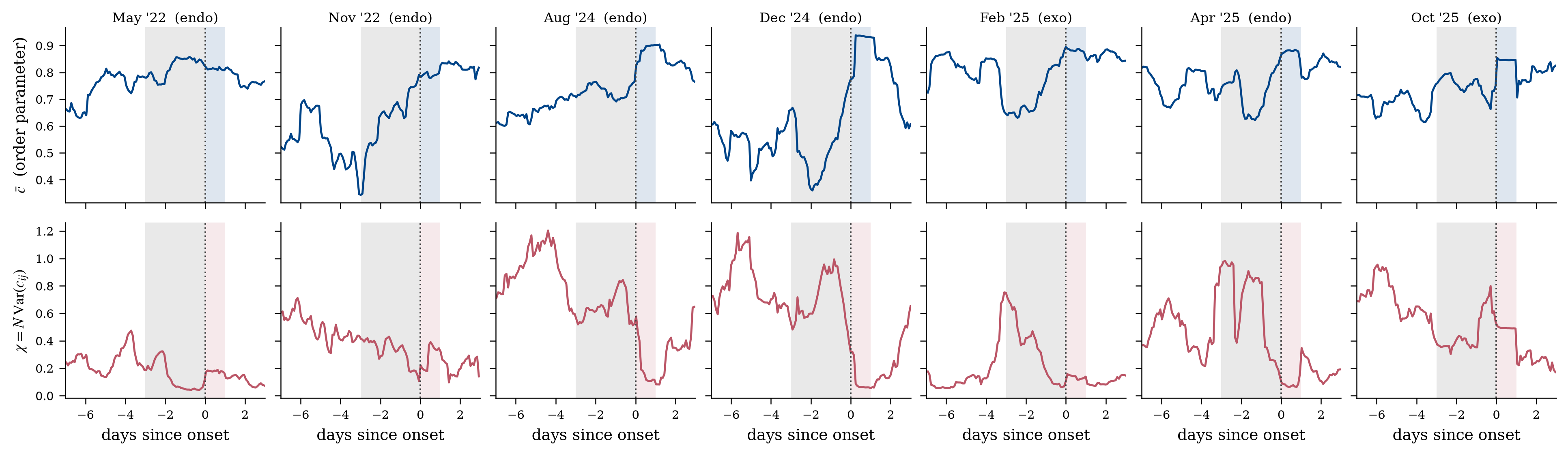}
  \caption{The fabric transition, event by event: order parameter $\bar c$
  (top row) and susceptibility proxy $\chi$ (bottom row) in one-day rolling
  windows stepped 2\,h, over $[-7, +3]$ days around onset (dotted line). At
  onset $\bar c$ jumps into the near-fully-ordered phase in six of seven
  events---May~2022, already ordered, is the grind exception---while $\chi$
  collapses rather than diverges; October~2025 is the outlier, de-correlating
  into onset with $\chi$ staying elevated. Shaded: the two windows the jump
  statistic compares---the three-day pre-onset mean (grey) against the first
  post-onset day (tinted). The load-bearing feature is the \emph{step} between
  them, not the drift within the pre-onset window; as Table~\ref{tab:fabric}
  reports, most pre-onset trends do not survive the overlap-aware null.
  Source: EXP-012.}
  \label{fig:fabric}
\end{figure}

\begin{figure}[t]
  \centering
  \includegraphics[width=\linewidth]{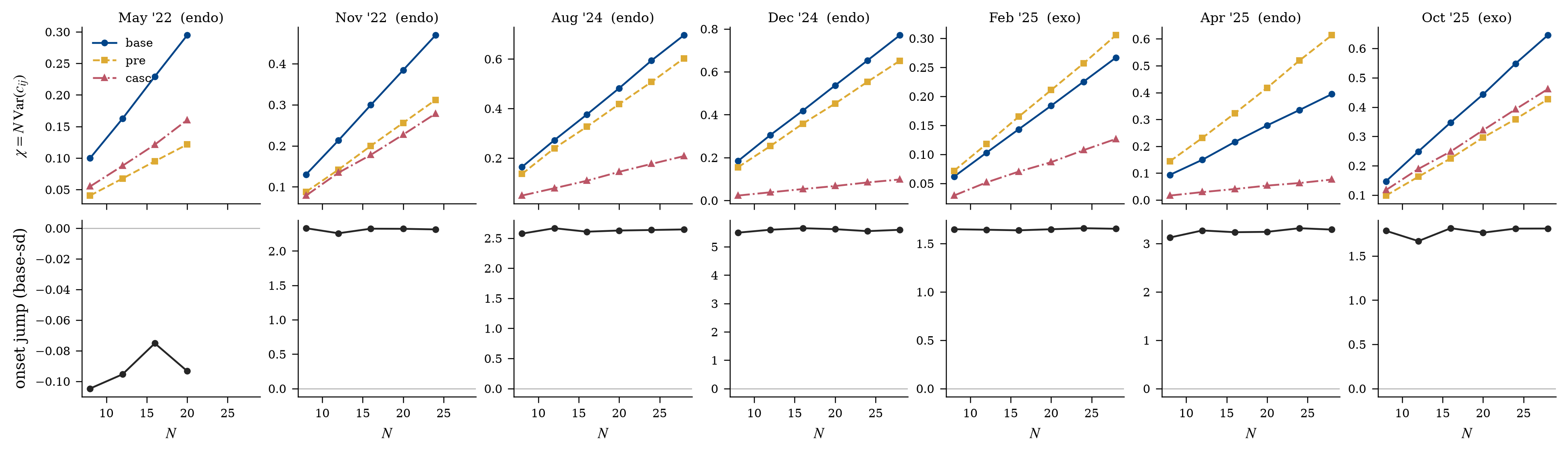}
  \caption{Finite-size behavior under subsampling ($N = 8$--$28$, 40 draws per
  size): $\chi$ is extensive in every regime (top; base / pre / cascade), and
  the onset jump in base-regime sd units is $N$-invariant (bottom). No critical
  scaling appears at any size. Source: EXP-016.}
  \label{fig:finitesize}
\end{figure}

% ---------------------------------------------------------------------
\section{The liquidity sector carries the universal signature}
\label{sec:liquidity}
The in-cascade signature is universal across venues and instruments, and it lives
in the price of liquidity---a market-liquidity spiral in the sense of
\citet{brunnermeier2009market}, and the order-book dry-up mechanism documented for
Bitcoin by \citet{donier2015}. On Binance, a Kyle-style impact regression
\citep{kyle1985} gives a
coefficient $k$ that spikes during every one of the seven cascades (six-hour means
elevated $\times 1.2$--$3.5$), with only a modest pre-onset build. On Hyperliquid
the same quantity is available \emph{directly} from quoted impact prices, with no
regression: the cascade impact-spread spikes for all six major instruments, by a
factor of $3.2$ to $9.1$ in the mean and $511$ to $927$ at the cascade-day
maximum, with a positive pre-onset trend in all six ($\tau = +0.06$ to $+0.24$).
Open interest clears hard---to $0.30$--$0.55$ of baseline on Hyperliquid ($-45$ to
$-70\%$), against $-24.6\%$ for Binance BTC. That two venues, two margining
conventions, and two independent measurements of impact agree that impact spikes
in every cascade---$\times 1.2$--$3.5$ where regressed, $\times 3.2$--$9.1$ where
quoted directly---is the universal fact any mechanism must reproduce.
\begin{figure}[t]
  \centering
  \includegraphics[width=\linewidth]{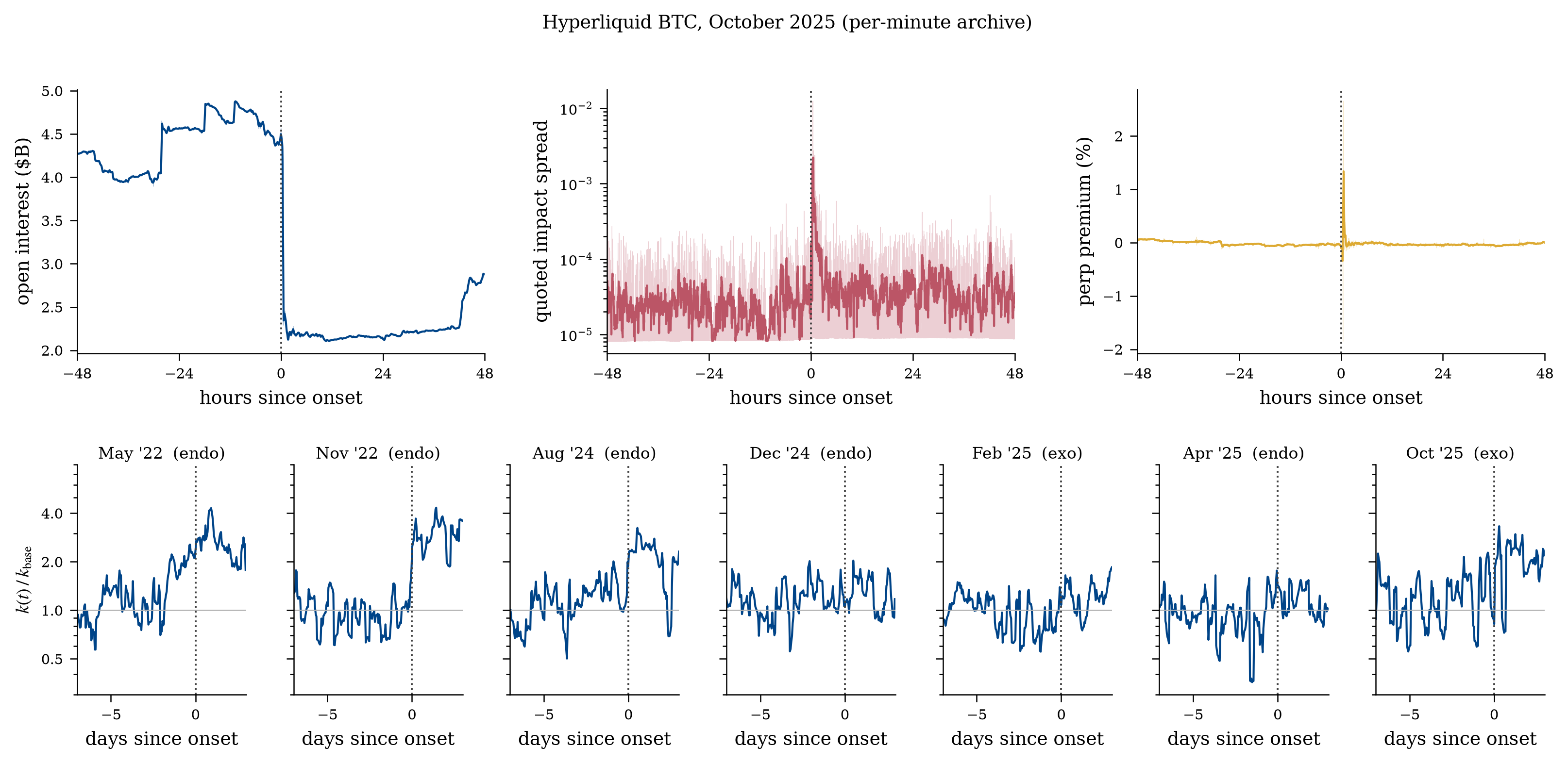}
  \caption{The in-cascade liquidity signature, on two venues and two
  instruments. \emph{Top:} Hyperliquid BTC through the October~2025 cascade
  (per-minute archive)---open interest, quoted impact spread (log scale) and
  perp premium around onset (dotted line); faint traces are per-minute values,
  solid lines a 15-minute rolling mean. \emph{Bottom:} the regressed Binance
  Kyle impact $k(t)$ for all seven events, each normalized by its own baseline
  median ($[-10,-3)$~d), log scale. The impact spike and the deep OI clearing
  are the universal in-cascade signature; the pre-onset build is modest and
  heterogeneous. Sources: EXP-017a, EXP-017b.}
  \label{fig:fragility}
\end{figure}

% ---------------------------------------------------------------------
\section{The branching model and its elimination}
\label{sec:lambda}
The natural mechanistic account is a Galton--Watson branching cascade. Forced
selling of notional $V$ moves the price by $\mathrm{d}p/p = k\,V$, with $k$ the
impact per dollar of net aggressor flow (units $\$^{-1}$); a relative move of size
$\delta$ sweeps the liquidation-threshold density $\rho(p)$ over an interval of
width $p\,\delta$, forcing a further $\rho(p)\,p\,\delta$ of notional. The
offspring mean of the resulting branching process is therefore
\begin{equation}
  \lambda = k\,\rho(p)\,p ,
  \label{eq:lambda}
\end{equation}
which is dimensionless. Here $p$ is the \emph{price level}, not a free parameter:
it appears because $\rho$ is a density per unit price while impact is expressed as
a relative move. Writing $\tilde\rho \equiv \rho(p)\,p$ for the forced notional per
unit \emph{relative} move---the form in which we measure it throughout---the ratio
is simply $\lambda = k\,\tilde\rho$, a product of two measurable quantities with no
free constant. The cascade condition is $\lambda \to 1$, and the amplification of
an initial shock $V_0$ is $V_0/(1-\lambda)$. The model makes two falsifiable
predictions; both fail.

The first is a timing prediction: if the pre-onset rise in $\lambda$ reflects a
genuine approach to the boundary, it should stand out against declines of
comparable depth that did \emph{not} cascade. It does not. Against placebo onsets
matched on the trailing three-day return, only four of seven events sit above the
matched median (Fisher $p = 0.062$; $p = 0.071$ for $\Delta\lambda$)---not
significant---and the effect is substantially mechanical: on placebos, $\lambda$
tracks the trailing price path itself
($\mathrm{corr}(\lambda_{\mathrm{pre}}, r_{3\mathrm{d}}) = -0.38$: the deeper the
trailing decline, the higher $\lambda$ reads). The lone exception is
February~2025 (matched percentile $1.00$, $z = +4.25$), recorded as a case
finding rather than a rule.

The second is a severity prediction: the zero-parameter unit slope relating
realized amplification to $-\log(1-\lambda)$. It is rejected in both units
(Table~\ref{tab:severity}),
$-0.244$ $[-0.767, +0.279]$ price-conditioned ($n = 157$) and $-0.152$
$[-0.703, +0.400]$ volume-native ($n = 158$), while the simulated power to detect
a true unit slope is $0.958$ and $0.970$ and the estimator is unbiased---so the
null is informative, not merely underpowered. The reason is structural: the
product $k\rho$ cancels itself, with $\mathrm{corr}(\log k, \log \rho) = -0.72$ and
an incremental $R^2$ of $\rho$ over $k$ that is zero to four decimal places.

The refutation hardens on directly measured impact. Replacing the proxied $k$ with
Hyperliquid quoted impact prices across five events, the unit slope is rejected at
every scale with power $\ge 0.99$ under two designs; the Binance proxy was itself
reliable ($\mathrm{corr} = +0.678$, above the $0.40$ attenuation boundary), so the
proxied null was not an artifact of measurement error. One sign result does
\emph{not} survive the upgrade, and we flag it as a methods lesson: the negative
loading $\beta_k < 0$ seen on proxied $k$ vanishes on measured $k$ ($+0.638$, not
significant); it sat entirely on the Kyle-estimator component orthogonal to true
impact. The punchline is uniform across proxied and measured regressors: none of the
scalar pre-state measures we can construct grades severity---not the stock of
leverage, not measured book fragility, not their product. This is a statement
about the quantities we can build from public pre-cascade data, not a proof that
no such quantity exists; Sec.~\ref{sec:discussion} names the one candidate that
our designs are structurally unable to test.
\begin{figure}[t]
  \centering
  \includegraphics[width=\linewidth]{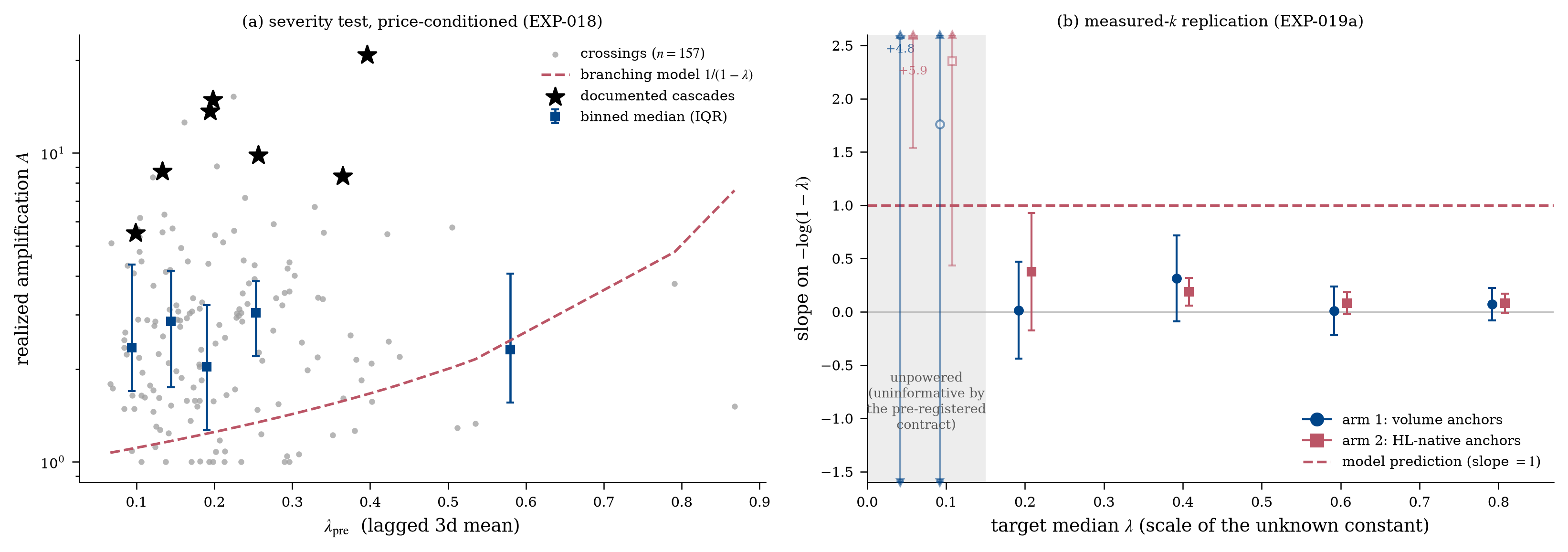}
  \caption{The severity test (price-conditioned design, $n = 157$): realized
  amplification $A$ against $\lambda_{\mathrm{pre}}$, with the zero-parameter
  branching prediction $A \propto 1/(1-\lambda)$ (curve), binned medians (IQR),
  and the seven documented cascades (stars); the binned medians are flat where
  the model requires them to rise. \emph{(b)} The same test with $k$ measured
  from Hyperliquid quoted impact prices, over the grid of scales for the
  unknown constant $c$ in $\lambda = c\,k\rho$ (EXP-019a), both anchor designs.
  Filled markers are powered ($\ge 0.99$); open markers in the shaded band are
  uninformative by the pre-registered contract. Sources: EXP-018, EXP-019a.}
  \label{fig:severity}
\end{figure}

\begin{table}[t]
  \centering
  \caption{The severity prediction across designs, units and venues. The
  branching model requires a slope of $1.0$ on $-\log(1-\lambda)$ and requires
  $\beta_k = \beta_\rho$. Power is the simulated probability of rejecting
  slope $= 0$ under a true unit-slope model at the observed $\lambda$ range and
  residual scale; the measured-$k$ rows are quoted at the target-median
  $\lambda$ nearest the proxied designs. All regressions include
  $\log V_0$ (or $\log X$), the trailing 3-day return, log realized volatility
  and event fixed effects. Sources: EXP-018, EXP-018c, EXP-019a.}
  \label{tab:severity}
  \begin{tabular}{llccccc}
    \toprule
    design & $k$ from & $n$ & slope on $-\log(1-\lambda)$ & power &
    $\beta_k$ & $\beta_\rho$ \\
    \midrule
    price-conditioned & Kyle regression & $157$ & $-0.244$ $[-0.767,+0.279]$ & $0.958$ & $-0.528$ & $+0.009$ \\
    volume-native     & Kyle regression & $158$ & $-0.152$ $[-0.703,+0.400]$ & $0.970$ & $-0.720$ & $+0.078$ \\
    volume anchors    & HL impact px    & $115$ & $+0.015$ $[-0.441,+0.471]$ & $0.995$ & $+0.638$ & $+0.222$ \\
    HL-native anchors & HL impact px    & $188$ & $+0.376$ $[-0.175,+0.928]$ & $1.000$ & $+0.090$ & $+0.420$ \\
    \bottomrule
  \end{tabular}
\end{table}

% ---------------------------------------------------------------------
\section{The engine, measured in flight}
\label{sec:engine}
Finally we measure the branching ratio of the record October~2025 cascade
\emph{in flight}, from the Hyperliquid fill log. Three complementary
constructions agree. They are not statistically independent---all three read the
same event, the same fill log and the same price path, so they share any common
timing or classification bias---but they rest on different assumptions, and a
shared artefact would have to survive all three to produce the agreement. The structural ratio $\hat\lambda_{\mathrm{struct}} = \hat k\,\hat{\tilde\rho}$,
with both factors measured in the units of Eq.~\eqref{eq:lambda} and no free
constants, traces $0.031$ (baseline) $\to 0.097$ (late pre-onset) $\to 0.195$
(nucleation) $\to 0.140$ (peak) $\to 0.032$ (late)---subcritical throughout, with
the full regime-by-regime decomposition in Table~\ref{tab:engine}. A
flow-based estimator (an INAR/Hawkes read of one-minute forced-sell counts, whose
calm-market level sits near $0.56$ from mechanical clustering and must be read as a
trajectory, not a level \citep{hardiman2013critical,kirchner2017estimation})
\emph{falls} to $0.28$ through nucleation rather than rising. And a direct
amplification bookkeeping---nucleation-window forced notional $V_0 = \$644$M
against a full-cascade total $V_{\text{total}} = \$733$M over the following
$15.7$~h---gives $A = 1.14$ and an implied $\lambda = 0.122$. All three place the
record cascade firmly below the critical boundary.

The cascade was not a slow chain reaction but an exogenous front-loaded sweep:
$87.8\%$ of all post-onset forced selling occurred in the first 30 minutes
($96.5\%$ within an hour), sweeping \$733M of book-hitting forced sales across
fifty-eight $0.25\%$-wide price buckets. Over the full Oct-9 to Oct-11 window the
same population sweeps \$0.96B across eighty buckets spanning
\$100.7k--\$122.7k, the additional \$225M having been swept \emph{before} onset;
the two totals are the same fills under the same sign and deduplication
conventions, differing only in the time window. Crucially, most of the offspring never touched
the order book. The venue backstop absorbed $62.6\%$ of post-onset forced-sell
notional off-book; in the worst minute (21:19~UTC) $\$641$M was force-sold, of
which the backstop took $\$576$M against only $\$64$M on the book. The branching
ratio is, in effect, \emph{engineered} down at the climax---an instance of the
automated-deleveraging trilemma \citep{chitra2025adl}---which yields a concrete,
falsifiable design prediction: venues without such a backstop vault should run a
hotter realized $\lambda$.
\begin{figure}[t]
  \centering
  \includegraphics[width=\linewidth]{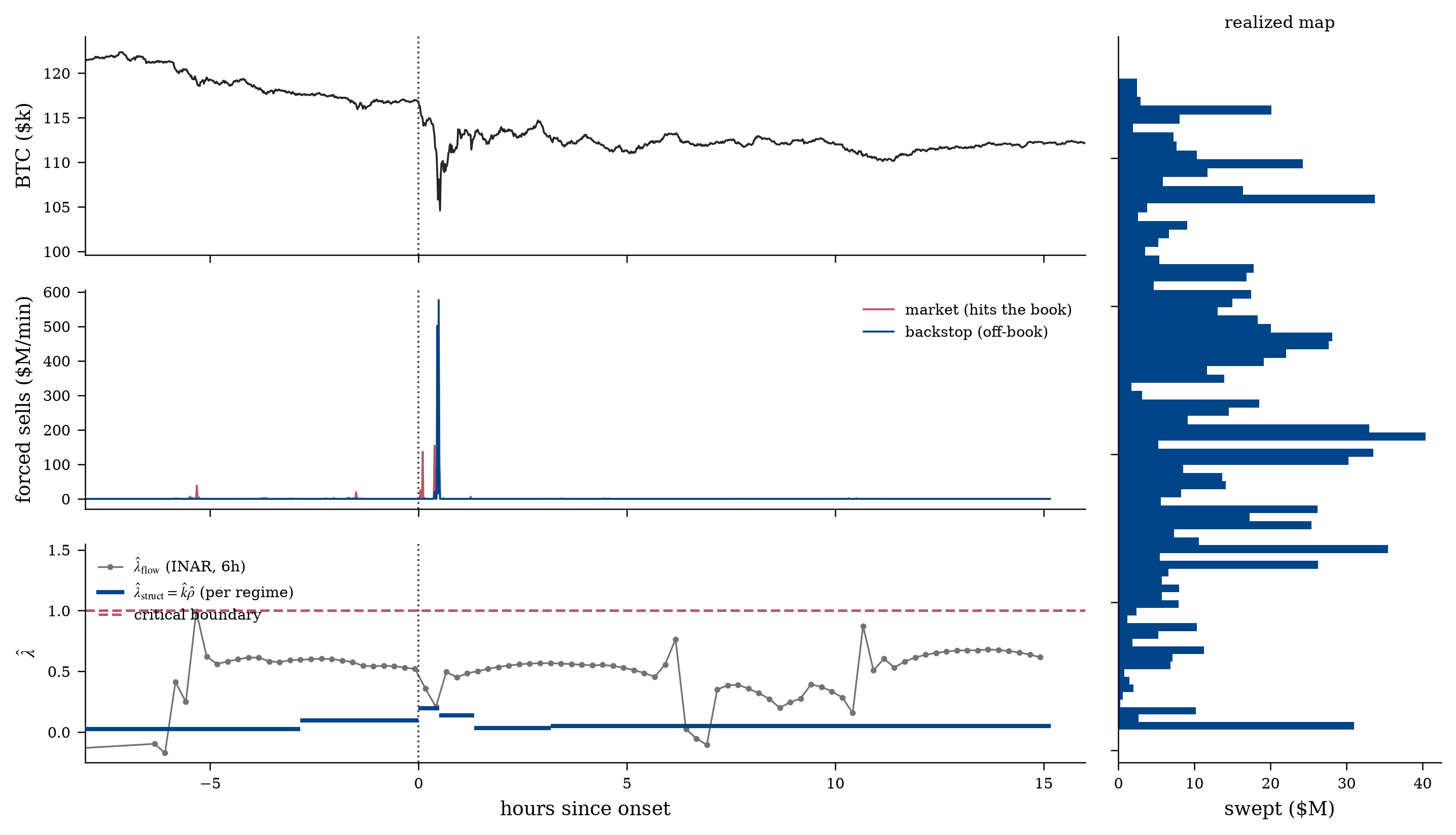}
  \caption{The October~2025 engine in flight, from the Hyperliquid fill log.
  \emph{Top to bottom:} the BTC mid price; measured forced sells per minute,
  split into the market leg that hits the book and the backstop leg absorbed
  off-book; and the two branching-ratio estimates---the flow-based
  $\hat\lambda_{\mathrm{flow}}$ (INAR on one-minute forced-sell counts) and the
  structural $\hat\lambda_{\mathrm{struct}} = \hat k\hat\rho$ per regime
  window---against the critical boundary $\lambda = 1$. \emph{Right:} the
  realized liquidation map over the full Oct-9 to Oct-11 window---\$0.96B of
  book-hitting forced sales across eighty $0.25\%$-wide buckets, of which \$733M
  landed after onset---sharing the price axis with the path that swept it.
  Source: EXP-019b.}
  \label{fig:engine}
\end{figure}

\begin{table}[t]
  \centering
  \caption{The engine by regime window (October~2025, Hyperliquid). Both
  factors of $\hat\lambda_{\mathrm{struct}}$ are measured from the fill log
  with no free constants: $\hat k$ is minus the slope of returns on net forced
  dollars, $\hat\rho$ the forced dollars per unit realized downmove.
  $\hat\lambda_{\mathrm{flow}}$ is the median of the INAR estimator on
  one-minute forced-sell counts; its calm-market level near $0.56$ is
  mechanical clustering, so the trajectory rather than the level carries the
  content \citep{hardiman2013critical,kirchner2017estimation}. The structural
  ratio stays below $0.2$ in every regime window, and the flow estimator's
  regime medians never exceed their calm-market level; individual six-hour
  windows of the flow estimator do exceed unity in the calm baseline
  ($\max 1.63$ coarse, $2.05$ notional-weighted), which is the mechanical
  small-sample behaviour the caveat above refers to and not an approach to
  criticality. Source: EXP-019b.}
  \label{tab:engine}
  \begin{tabular}{lccccc}
    \toprule
    window (UTC) & $\hat k$ [\$$^{-1}$] & $\hat\rho$ [\$] &
    $\hat\lambda_{\mathrm{struct}}$ & $\hat\lambda_{\mathrm{flow}}$ &
    forced \\
    \midrule
    baseline (Oct-9)          & $1.01\times10^{-10}$ & $3.11\times10^{8}$ & $0.031$ & $0.566$ & \$80M \\
    pre (Oct-10 00--18)       & $6.39\times10^{-11}$ & $4.09\times10^{8}$ & $0.026$ & $0.569$ & \$91M \\
    late-pre (18--20:50)      & $9.86\times10^{-11}$ & $9.80\times10^{8}$ & $0.097$ & $0.561$ & \$55M \\
    nucleation (20:50--21:20) & $3.33\times10^{-11}$ & $5.87\times10^{9}$ & $0.195$ & $0.283$ & \$644M \\
    peak (21:20--22:10)       & $2.06\times10^{-10}$ & $6.76\times10^{8}$ & $0.140$ & $0.484$ & \$71M \\
    late-casc (22:10--24:00)  & $2.38\times10^{-9}$  & $1.34\times10^{7}$ & $0.032$ & $0.548$ & \$1M \\
    aftermath (Oct-11 am)     & $6.31\times10^{-10}$ & $7.99\times10^{7}$ & $0.050$ & $0.547$ & \$18M \\
    \bottomrule
  \end{tabular}
\end{table}

% ---------------------------------------------------------------------
\section{Discussion}
\label{sec:discussion}
These four results are one story. Severity is set by
shock $\times$ map-in-path $\times$ liquidity withdrawal: a first-order jump in the
correlation fabric, a subcritical engine that clears its fuel almost instantly, and
a pre-cascade state that grades nothing, together explain why Part~I found no
reliable early-warning signal---the EWS failures are the single-series shadow of a
transition that is collective and abrupt rather than gradual and critical.

The central boundary is the venue scope of $\lambda$. Our measured branching ratio
is \emph{within-venue}; the system-level feedback loop runs through the price shared
across venues, whose transparent, mirrored coupling we documented separately.
Within-venue subcriticality does not preclude system-level amplification, and the
network version of the branching measurement is left to future work.

The backstop result speaks to mechanism design: the vault acts as a circuit breaker
on branching, trading against the automated-deleveraging trilemma of
\citet{chitra2025adl}. The contrast case is instructive---Binance's \$283M
reimbursement and its pre-oracle collateral pricing during the same event
\citep{amberdata2025oct} are what the absence of that circuit breaker looks like.

Finally, the results define what a mechanistic model must reproduce, a falsifiable
target list for Part~III: (a) a first-order order-parameter jump with $\chi$
collapsing; (b) impact spiking $\times 3$--$9$ in-cascade; (c) a subcritical
in-flight $\lambda$ with front-loaded forced flow; and (d) no single-variable
early-warning signal yet a fabric-level build-up, as in the February~2025
signature. One pre-state object remains untested by this program: the accumulated
map of liquidation thresholds near price---measurable from the fill log---which
is not a scalar but a measure swept by the path, and is therefore exactly the
kind of state variable our scalar severity designs cannot rule out, and the
natural place for a sequel to begin.

A further boundary is the aggregation itself. Equation~\eqref{eq:lambda} is a
local, differential statement in $(p,t)$, whereas $\hat k$ and $\hat{\tilde\rho}$
are window estimates over regimes in which price moves by several percent and the
book is visibly non-stationary. The regime windows are chosen short enough that
the two factors are not averaged across the nucleation boundary, but the product
of two window means is not the window mean of the product, and no part of our
argument should be read as identifying an instantaneous offspring mean.

Limitations: seven events; BTC-centric severity designs; open-interest-proxied
$\rho$ on the historical events, since the fill archive begins 2025-05-25; a single
case-study event for the in-flight measurement; and a backstop that actively
suppresses the very quantity we measure on the one venue where we can measure it.

% ---------------------------------------------------------------------
\section{Conclusion}
\label{sec:conclusion}
Across seven major crypto-perpetual liquidation cascades, the transition is
first-order and its signature lives in the liquidity sector, not in the memory of
any single price. None of the scalar pre-state measures we were able to
construct grades severity---leverage stock, measured book fragility, or their
product---while the one candidate our designs cannot reach, the accumulated map
of liquidation thresholds along the path, remains open:
leverage is the fuel, not the fire alarm, and the fire is in the book. On the one
venue where the engine can be watched directly, it never went critical---the record
cascade of October~2025 ran subcritical, front-loaded, and backstop-absorbed.
Whether that subcriticality survives the cross-venue loop is the question we
would put next.

% ---------------------------------------------------------------------
\appendix
\section{Spurious dark causality from timestamp misalignment}
\label{app:spurious}
A cross-venue analysis of the October~2025 event by Pattern Causality yields a
methodological result worth isolating, because it can silently corrupt any
mixed-source high-frequency study. Binance \emph{metrics} dumps stamp the interval
\emph{end}, while kline timestamps mark the interval \emph{start} (empirically, the
metric taker ratio and the kline-derived ratio correlate at $0.996$ at exactly
$+5$~min, versus $\le 0.11$ at every other lag). Pairing the two conventions
naively---flow one bar ahead of returns---converts the transparent, mirrored
leverage--price coupling into an apparent spectrum that is roughly
zero-positive, half-negative, and half-\emph{dark} at twice the surrogate null,
including a statistically significant cascade-week ``dark causality'' rise
(Mann--Whitney $p < 10^{-4}$) that evaporates entirely under correct alignment.
Cross-mapping spectra are
alignment-fragile; mixed-source financial pipelines---the principal use case of
Pattern Causality---require a timestamp-semantics audit before any spectral claim.
This exhibit also cautions linear-Granger network studies built on mixed feeds,
such as \citet{shukla2026networks}.

\section*{Reproducibility}
Every figure and number in this paper is produced by a script from a frozen
experiment record, cited inline as \texttt{EXP-012}--\texttt{EXP-019b}; the
records fix the question, data, method, numbers and verdict at run time and are
append-only. All three data sources are public and reconstructible without
proprietary feeds: Binance panel and BTC series from the public
\texttt{data.binance.vision} dumps; the Hyperliquid per-minute asset-context
archive from \texttt{s3://hyperliquid-archive} (requester-pays); and the
Hyperliquid node fill log from \texttt{s3://hl-mainnet-node-data}. Two
conventions are load-bearing for anyone reproducing the numbers and are stated
where they are used: Binance metrics dumps stamp the interval end while klines
stamp its start (a $-5$~min realignment, Appendix~\ref{app:spurious}), and the
fill log's liquidation object rides both legs of a forced trade, so totals must
be deduplicated by the liquidated-user leg. Analysis code is available from the
author on request.

\bibliographystyle{plainnat}
\bibliography{refs}

\end{document}